\documentclass[doublecol]{epl2}
\usepackage{slashbox}
\usepackage{graphicx}
\usepackage{amsmath}
\newcommand{\up}{\uparrow}
\newcommand{\dn}{\downarrow}
\newcommand{\ba}{\begin{eqnarray}}
\newcommand{\ea}{\end{eqnarray}}

\newcommand{\be}{\begin{equation}}
\newcommand{\ee}{\end{equation}}
\newcommand{\la}{\langle}

\newcommand{\ra}{\rangle}
\newcommand{\ete}{{\it et al.}}
\newcommand{\et}{{\it et al. }}

\def\prl{{\it Phys. Rev. Lett.}}

\def\prb{{\it Phys. Rev. B}}

\def\cpl{{\it Chem.\ Phys.\ Lett.\ }}

\def\jpcm{{\it J. Phys.: Condens. Matter}}

\begin{document}

\title{Laser-induced spin {protection} and switching in a
  specially designed {magnetic dot}: A theoretical
  investigation}

\author{G. P. Zhang\inst{1}\thanks{}, M. S. Si\inst{1} and T.F. George\inst{2}}

\institute{
\inst{1} Department of Physics, Indiana State University, Terre Haute, Indiana 47809, USA\\
\inst{2} Office of the Chancellor and Center for Nanoscience, Departments of
Chemistry \& Biochemistry and Physics \& Astronomy, University of Missouri-St. Louis,
St.  Louis, MO 63121 USA }

\abstract{Most laser-induced femtosecond magnetism investigations
  are done in magnetic thin films. Nanostructured {magnetic
    dots}, with their reduced dimensionality, present new
  opportunities for spin manipulation. Here we predict that if a
  {magnetic dot} has a dipole-forbidden transition between
  the lowest occupied molecular orbital (LUMO) and the highest
  unoccupied molecular orbital (HOMO), but a dipole-allowed transition
  between LUMO+1 and HOMO, {electromagnetically induced
    transparency can be used to prevent ultrafast laser-induced spin
    momentum reduction, or spin protection.}  This is realized through
  a strong dump pulse to funnel the population into LUMO+1.  If the
  time delay between the pump and dump pulses is longer than 60 fs, a
  population inversion starts and spin switching is achieved.  These
  predictions are detectable experimentally.}

 \maketitle

\begin{table}
\caption{ Eigenvalues, and spin and orbital momenta of fcc Ni at
  $k$-point of $k=(101,69,-7)/104$.  All the results are computed
  using the first-principles method.  The Fermi energy is at 0 eV.
  The $k$-mesh is $104\times 104 \times 104$, so all the numbers are
  divided by 104. }
\begin{center}
\begin{tabular}{cccc}
\hline
\hline
 $i$ & ${\cal E}_i$ (eV) & ~~$S_z(\hbar)$~~ &~~ $L_z(\hbar)$\\
\hline
     1&    -4.065&     0.499&    -0.009\\
     2&    -3.740&     0.496&    -0.049\\
     3&    -3.691&    -0.494&     0.003\\
     4&    -3.112&    -0.498&     0.054\\
     5&    -2.311&     0.496&    -0.024\\
     6&    -1.682&    -0.489&     0.019\\
     7&    -1.109&     0.490&     0.052\\
     8&    -0.565&     0.174&    -0.028\\
     9&    -0.499&    -0.174&    -0.001\\
    10&     0.351&    -0.499&    -0.017\\
    11&     4.863&     0.500&     0.002\\
    12&     5.086&    -0.500&    -0.003\\
    13&     9.873&     0.498&    -0.008\\
    14&    10.040&    -0.487&     0.007\\
    15&    10.261&     0.487&     0.009\\
    16&    10.434&    -0.498&    -0.008\\
    17&    12.152&     0.500&     0.001\\
    18&    12.252&    -0.500&    -0.001\\
\hline
\hline
\end{tabular}
\end{center}
\label{tablefccni}
\end{table}

\begin{table}
\caption{Eigenvalues (${\cal E}_i$) and their spin ($S_z$) and
  orbital momenta ($L_z$).  The spin-orbit coupling is chosen as  $\lambda=0.05$
  eV.  The Fermi energy is set between $d_5$ and $d_6$ in order to
  have small orbital and large spin momenta.  }
\begin{center}
\begin{tabular}{ccccc}
\hline
\hline
 $i$ &Orbital type& ${\cal E}_i$ (eV) & ~~$S_z(\hbar)$~~ &~~ $L_z(\hbar)$\\
\hline
   1&$d_1 $&  -2.496&   ~0.444&   -0.072\\
   2&$d_2 $&  -2.271&   -0.443&   ~0.073\\
   3&$d_3 $&  -1.754&   ~0.479&   ~0.008\\
   4&$d_4 $&  -1.547&   -0.479&   -0.009\\
   5&$d_5 $&  -0.750&   ~0.497&   -0.003\\
   6&$d_6 $&  -0.550&   -0.497&   ~0.003\\
   7&$p_1 $&  ~0.015&   ~0.350&   -0.013\\
   8&$p_2 $&  ~0.086&   -0.350&   ~0.013\\
   9&$d_7 $&  ~0.248&   ~0.487&   -0.016\\
  10&$d_8 $&  ~0.452&   -0.486&   ~0.017\\
  11&$d_9 $&  ~0.973&   ~0.448&   ~0.068\\
  12&$d_{10} $&  ~1.195&   -0.450&   -0.069\\
  13&$p_3 $&  ~1.440&   ~0.499&   -0.001\\
  14&$p_4 $&  ~1.490&   -0.499&   ~0.001\\
  15&$p_5 $&  ~2.845&   ~0.357&   ~0.012\\
  16&$p_6 $&  ~2.915&   -0.357&   -0.012\\
\hline
\hline
\end{tabular}
\end{center}
\label{table}
\end{table}

\begin{figure*}[htb]
\onefigure[angle=0,width=8cm]{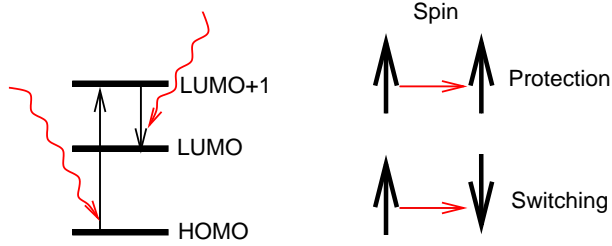}
\caption{
Laser-induced spin {protection} and spin switching. A three-level
system is shown. The transition between LUMO and HOMO is forbidden.}
\label{fig0}
\end{figure*}

\begin{figure*}[htb]
\onefigure[angle=270,width=14cm]{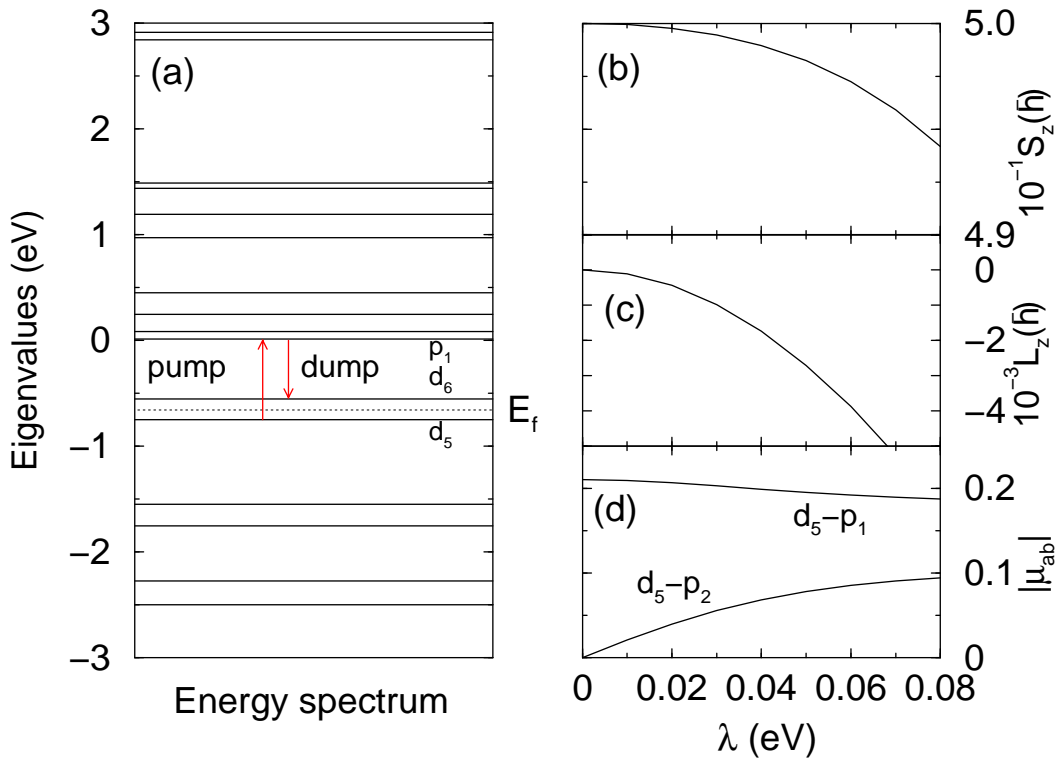}
\caption{(a) Energy spectrum.  The results are obtained with
  $\lambda=0.05$ eV.  $E_f$ represents the Fermi energy.  The pump
  laser is tuned to the transition between $d_5$ and $p_1$, while the
  dump to the transition between $d_6$ and $p_1$.  (b) Ground-state
  spin angular momentum change as a function of $\lambda$. (c)
  Ground-state orbital angular momentum change with $\lambda$. (d)
  Transition matrix elements $|\mu_{ab}|$ between $d_5$ and $p_1$ and
  between $d_5$ and $p_2$ as a function of $\lambda$.  }
\label{fig1}
\end{figure*}

\begin{figure*}[htb]
\onefigure[angle=270, width=14cm]{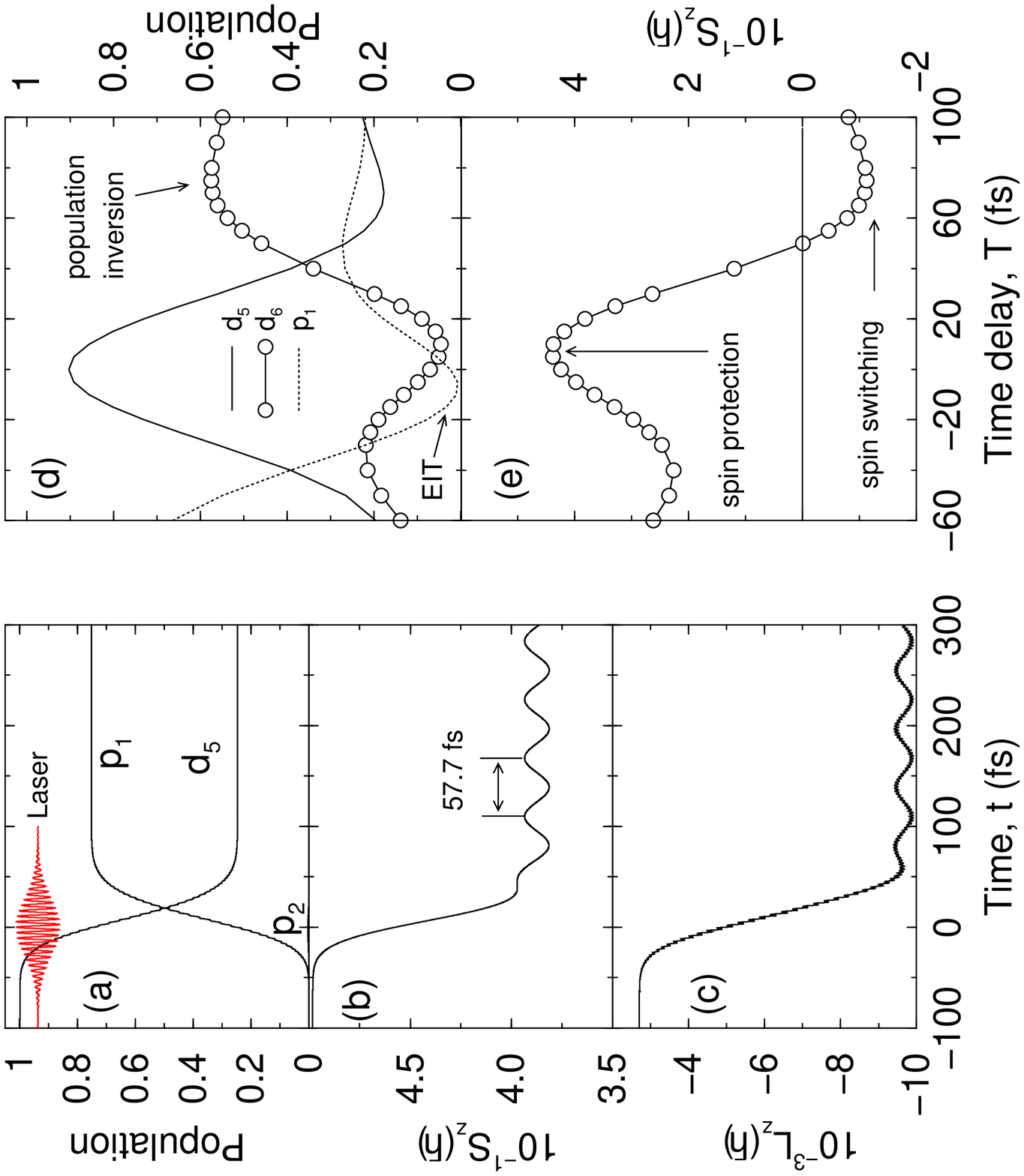}
\caption{(a) Population change as a function of time upon laser
  excitation. Only the three states $p_1$, $p_2$ and $d_5$ are shown
  (the population in all the other states is nearly unchanged).
  Inset: Laser pulse.  (b) Spin momentum change as function of
  time. The oscillation is due to the interference between states
  $p_1$ and $p_2$. (c) Orbital angular momentum change.  (d)
  Population change excited by pump and dump pulses as a function of
  the time delay between the two pulses. Electromagnetically induced
  transparency (EIT) occurs for the transition between $d_5$ and $p_1$
  when the delay is close to 0 fs, and population inversion occurs for
  the delay longer than 60 fs. (e) Change from spin
  {protection} to spin switching, where the time delay
  between the pump and dump pulses is longer than 60 fs.
\label{fig2}}
\end{figure*}

\section{Introduction}

The discovery of femtosecond demagnetization in ferromagnetic Ni \cite{beaurepaire}, which has attracted tremendous attention
worldwide \cite{koopmans,new,naturephysics09, bigot01, prl09, jpcm10},
challenges traditional wisdom and opens a new technological frontier
for fast magnetic switching. The nonthermal inverse Faraday effect
\cite{stanciu} is an excellent example, where one can manipulate the
spin via the polarization of light, without involvement of the
lattice.  This has further inspired new theoretical and experimental
investigations into the mechanism of the femtosecond magnetism.  Up to
now, while the experimental focus has been on thin films and bulk
materials \cite{prl00,malik}, magnetic nanostructures, with reduced
dimensionality, unique symmetry and quantum confinement, are very
attractive. For instance, their electronic and magnetic properties can
be tailored systematically.  Experimental studies in cobalt
nanostructures show very pronounced spin excitations \cite{bigotco}.
This presents a unique opportunity for electromagnetically induced
transparency (EIT) in a spin system.  EIT has been demonstrated in
atoms and molecules \cite{eitt}, and can be realized in some special
designed quantum dots, with a proper state symmetry.  EIT in a
three-level system relies on two coherent laser pulses: one as a
control pulse and the other as a probe pulse. The absorption of the
probe depends on the control pulse (see fig. \ref{fig0}).  It would be
fantastic if such EIT can be realized in a {magnetic dot},
since one can then fully integrate the light control into the magnetic
storage on a femtosecond time scale, a daunting but technologically
significant task.  This motivates us to pursue such a system.

In this paper, we show that {in a carefully designed
  magnetic dot, EIT can be used to protect spin momentum reduction.}
Using two laser pulses, one pump and one dump, we can not only protect
a spin state but also switch it.  The former is done through EIT, and
the latter is achieved through population inversion. {Spin
  protection} does not occur naturally in any {magnetic
  dots}.  The minimum condition for this to occur is that the optical
transition between the lowest unoccupied molecular orbital (LUMO) and
the highest occupied molecular orbital (HOMO) must be dipole-forbidden,
while the transition between HOMO and LUMO+1 or higher states is
dipole-allowed. The spin-orbit coupling must not be zero. Our
investigation is based on two different theoretical formalisms. We
start with the first-principles calculation in ferromagnetic fcc bulk
nickel, and we then construct a model system with nearly identical
electronic and magnetic properties. The model system allows us to
search the best candidate for {spin protection and
  switching}.  Future experiments can directly test our predictions.

This paper is arranged as follows. In the second section, we present our
theoretical scheme, followed by the laser-induced ultrafast
demagnetization in the third section. {The spin protection} and
switching is presented in the fourth section, and we conclude this paper in the fifth section.

\section{Theoretical formalism}

\newcommand{\ik}{i{\bf k}}
\newcommand{\jk}{j{\bf k}}

The first-principles method is certainly a method of choice for
quantum dots, but there is a limit on what the method can do. It is
difficult to isolate one physical quantity without affecting others in
question. For instance, the spin-orbit coupling is a joint effect of
the crystal potential and electron wavefunction, and can not be
parametrized easily at the first-principles level.  To strike a
balance between a real system and a model system, we first perform a
first-principles investigation in fcc nickel. While details have been
presented elsewhere \cite{prb09}, here in brief we use the
full-potential augmented plane wave method within the density
functional formalism as implemented in the WIEN2k code \cite{wien}. By
solving the Kohn-Sham equation (in Ry atomic units),

\be
[-\nabla^2+V_{Ne}+V_{ee}+V_{xc}^\sigma]\psi_{\ik}^\sigma(r)=E_{\ik}^\sigma
\psi_{\ik}^\sigma (r) \label{ks},
\ee
we obtain the eigenfunctions and eigenenergies, from which we construct the spin and orbital matrix
elements among all the band states. In eq. (\ref{ks}), the first and
second terms are kinetic energy and Coulomb interaction between the
electron and the nuclei, and $V_{xc}$ is the Coulomb and exchange
interactions. The spin-orbit coupling is included via the second-order
variational method.  In table \ref{tablefccni}, we show the
eigenenergies, spin and orbital matrix elements at one crystal
momentum point $k$. These elements are the basis for our model.

\begin{widetext}
\be
H_d=\sum_{m\sigma}\varepsilon_{d\sigma} c_{m\sigma}^\dag
c_{m\sigma} + \sum_{m_1,m_2;\sigma_1,\sigma_2} \lambda \la
2,m_1,\sigma_1|\vec{l}\cdot\vec{s}|2,m_2,\sigma_2\ra
c_{m_1\sigma_1}^\dag c_{m_2\sigma_2} +\sum_{m\sigma} \delta
c_{m+1\sigma}^\dag c_{m\sigma} +h.c. ~,
\label{ham}
\ee
\end{widetext}

We introduce a sixteen-level model, with the spin-orbit coupling as an
input parameter.  The model consists of three $p$-orbitals ($Y_{11}$,
$Y_{10}$ and $Y_{1-1}$) and five $d$-orbitals ($Y_{22}$, $Y_{21}$,
$Y_{20}$, $Y_{2-1}$, and $Y_{2-2}$) for each spin channel.  Here the
$Y_{lm}$'s are the spherical harmonics, and the radial part of the
wavefunction is included through the parameters (see below).  The
overlap between $p$- and $d$-orbitals is ignored to minimize the
number of parameters used, but the overlap between $d$-orbitals and
that between $p$-orbitals are introduced.  To mimic a ferromagnetic
ground state, these $d$-orbitals lie about 2 eV below the
$p$-orbitals.  Within these approximations, our Hamiltonian consists
of two parts, $H_0=H_d+H_p$.  Here the Hamiltonian for $d$-states is

\begin{center}
See eq. \ref{ham}
\end{center}
where $\varepsilon_{d\sigma}$ is the $d$-orbital energy with spin
index $\sigma$, $\vec{l}$ is the orbital angular momentum operator,
and $\vec{s}$ is the spin angular momentum operator.
$c_{m\sigma}^{\dagger}$ $(c_{m\sigma})$ is the electron creation
(annihilation) operator, creating (annihilating) an electron in
orbital $m$.  We choose the spin-orbit coupling (SOC) constant $\lambda=0.05$ eV, except in
figs. \ref{fig1}(b), \ref{fig1}(c) and \ref{fig1}(d).  The Dirac
bracket $|2,m,\sigma\ra$ in eq. (\ref{ham}) represents $Y_{2m}$
multiplied by the spin wavefunction $|\sigma\ra$. The last term,
originating from the crystal and exchange fields,
 denotes the overlap
between orbitals with different $m$, with $\delta=1$ eV. (A similar
Hamiltonian can be set up for the $p$-orbitals.)  Based on our
first-principles calculation for fcc Ni \cite{prl00,prb09}, we choose
$\varepsilon_{d\up}=-0.75 $ eV, $\varepsilon_{d\dn}=-0.55 $ eV,
$\varepsilon_{p\up}= 1.44 $ eV and $\varepsilon_{p\dn}= 1.49 $ eV.
While the precise orbital energies are not critical here, in order to
realize a strong spin momentum, we find it is necessary to adjust the
overlap term if different orbital energies are used.  An extension to
$f$-orbitals is straightforward.

We first compute the spin, orbital and dipole transition matrices by
diagonalizing the Hamiltonian.  The orbital characters of the
eigenstates are determined by their eigenvectors.  The eigenvalues,
spin and orbital momenta are tabulated in table \ref{table}.  The six
lowest eigenstates are all $d$-states, followed by two $p$-states.  It
is clear from the table that to realize large spin but small orbital
momenta, the number of occupied states is not arbitrary. We choose the
Fermi energy between $d_5$ and $d_6$ so the five lowest $d$-orbitals
are occupied. The resultant ground-state spin momentum is $\la
s_z\ra=0.498 \hbar$, and the orbital momentum is $\la l_z\ra= -0.0027
\hbar$.  This result can be compared with our first-principles results
in table \ref{tablefccni}, and we find that they match very well. For
instance, the majority of spin values are around $0.5\hbar$ and the
orbital momentum is below $0.1\hbar$.  This gives us confidence that
our model is able to simulate a ferromagnetic system.

 Figure \ref{fig1}(a) shows the energy spectrum with the relevant
 transition states labeled.  Figures \ref{fig1}(b) and \ref{fig1}(c)
 show the dependence of the spin ($s_z$) and orbital angular momenta
 ($l_z$) on the SOC constant $\lambda$. It is interesting to note that
 the spin decreases with $\lambda$, but we also find this to be
 parameter dependent, and in some cases the dependence is not
 monotonic. The orbital momentum is zero if $\lambda$ is zero, or
 complete quenching, but with nonzero $\lambda$ the quenching is not
 complete. This reproduces the well-known fact that the SOC drags some
 orbital momentum with it.  We find transition matrix elements are
 larger between two states with a similar spin moment, which
 independently validates our prior first-principles
 results \cite{prb09,prb08}. For instance, the transition matrix
 element ($z$-direction) between $d_5$ ($s_z=0.497\hbar$) and $p_1$
 ($s_z=0.350\hbar$) is 0.196 $\xi$, while the element between $d_5$
 ($s_z=0.497\hbar$) and $p_2$ ($s_z=-0.350\hbar$) is -0.078 $\xi$,
 where $\xi$ is a constant resulting from the radial part of the
 wavefunction. This demonstrates again that our model is applicable to
 ferromagnetic systems.  Therefore, even with SOC, the optical
 transition still prefers spin-conserved transitions over
 spin-nonconserved transitions, though for the demagnetization
 process, the latter transitions are most important. The systematic
 change of the transition matrix elements with the SOC for the above
 two transitions is shown in fig. \ref{fig1}(d). To our knowledge,
 this is the first numerical result that shows clearly the transition
 matrix element drops for a ``spin-conservation'' transition (here
 between $d_5$ and $p_1$), while the elements for the spin-flip
 transition (between $d_5$ and $p_1$) increase. This demonstrates the
 critical role of the SOC in ultrafast demagnetization.

\section{{Laser-induced coherent magnetization}}

To simulate the ultrafast spin evolution, we introduce a laser pulse
with $E_i(t)=A_i\cos(\omega_i t)\exp[-(t-t_i)^2/\tau_i^2]$, where
$A_i$, $\omega_i$, $t_i$ and $\tau_i$ are the laser amplitude,
frequency, time delay and pulse duration of pulse $i$, respectively,
and $t$ is time.  The laser has a Gaussian shape, with duration of 40
fs (see the inset of fig. \ref{fig2}(a)).  We tune the laser frequency
to be resonant with the dipole-allowed transition between $d_5$ and
$p_1$. The dynamic simulation starts with solving the Liouville
equation for the density matrix $\rho$ as \cite{naturephysics09,prb09,prl08},
\be i\hbar \frac{\partial
  \rho}{\partial t} =[H,\rho], \ee
where $H$ consists of the
Hamiltonians $H_d$ and $H_p$ for the $d$- and $p$-states and the
interaction $H_I$ between the system and laser
field. $H_I=-eE(t)\sum_{ij}\la i |z| j\ra c^\dag_ic_j$, where $-e$ is
the electron charge, the light polarization is along the
$z$-direction, and $\la i |z| j\ra$ is the transition matrix element
between states $i$ and $j$.  The merit of our approach should not be
underestimated. Although the Liouville equation is equivalent to the
time-dependent Schr\"odinger equation, the Liouville equation naturally
takes into account the Pauli exclusion principle and includes the
antisymmetry of the many-body wavefunction.  However, most of the
time-dependent Schr\"odinger equations from textbooks are for a
single particle only, which is not suitable even for a non-interacting
case.

The general form of the density matrix is a convoluted integration
equation.  We express the density matrix in a state representation as
\be i\hbar\frac{\partial \rho_{nm}}{\partial t}=({\cal E}_n-{\cal
  E}_m)\rho_{nm} +\sum_k (H^I_{nk}\rho_{km} - \rho_{nk}H^I_{km}), \ee
where $k$, $n$ and $m$ are the state indices, ${\cal E}_n$ is the
eigenenergy of state $n$, $\rho_{nm}$ is the density matrix between
states $n$ and $m$, and $H^I$ is the interaction between the laser
field and the system. If we integrate over time, we have

\begin{center}
See eq. \ref{eq5}
\end{center}
which is the general form of the density matrix.  It is clear that the time
evolution of the density depends on its history and the laser field.
Further simplifications are possible, only if we know the profile of
the laser field. For instance, if the laser is a continuous wave and
is very week, the density matrix can be approximated by the first
order density matrix whose time evolution is proportional to
$\exp[-i(\omega_{laser}-\omega_{nm})t] $, where $\omega_{nm}$ is the
transition frequency between states $n$ and $m$, $\omega_{laser}$ is
the laser frequency, and $t$ is the time. If the laser is a pulse,
then the profile of the density matrix sensitively depends on the
laser pulse duration, laser frequency and the electronic states
involved as can be seen from the above equation. In this case, an
analytic solution is generally not possible, and a numerical
integration is a must.

\begin{widetext}
\be
\rho_{nm}(t)=\frac{1}{i\hbar} \int^t_{-\infty} dt' ({\cal E}_n-{\cal
  E}_m)\rho_{nm}(t') +\frac{1}{i\hbar} \sum_k
\int^t_{-\infty} dt'
(H_{nk}^I(t')\rho_{km}(t)-\rho_{nk}(t')H^I_{km}(t') ),
\label{eq5}
\ee
\end{widetext}

 Figure \ref{fig2}(a) shows {that} the population change
 closely follows the profile of the laser field.  Upon the laser
 excitation, the $d_5$-state loses its population from 1 to 0.25
 within the first 40 fs. In the meantime, the $p_1$-state gains the
 same amount of population from 0 to 0.75. Other states have a very
 small change. For instance, $p_2$ only has a tiny increase around 0
 fs. Once the laser field is over, the population or diagonal element
 of the density matrix $\rho_{ii}$ change is stabilized. However, this
 does not mean that all the elements of the density matrix become time
 independent. The interference between states, which is induced by the
 laser field initially, remains. This is precisely what happens with
 the spin momentum change.

Figure \ref{fig2}(b) shows the spin change as a function of time. The
spin drops similarly within the first 40 fs, followed by an oscillation
with period 57.7 fs. The first drop results from the spin momentum
difference between $d_5$ and $p_1$, as seen from table \ref{table},
where $\la d_5 |s_z| d_5\ra =0.497 \hbar$ and $\la p_1 |s_z| p_1\ra
=0.350 \hbar$. The oscillation is from the interference between $p_1$
and $p_2$, which can be verified by switching on/off the density
matrix element $\rho_{p_1,p_2}$. It is remarkable that even though the
population of $p_2$ is almost zero, its interference with $p_1$ is
directly responsible for this salient change. The period 57.7 fs
corresponds to 0.071 eV, which matches the energy gap between $p_1$
and $p_2$ exactly. The orbital momentum change is shown in
fig. \ref{fig2}(c). In contrast to the spin change, the orbital
angular momentum increases (more negative). Besides its initial drop,
the orbital momentum also has smaller rapid beatings.

\section{{Spin protection} and switching}

As mentioned in the Introduction, the key to the electromagnetically
induced transparency relies on the state symmetry.  In the following,
we demonstrate that one can not only reduce and flip a spin (spin
switching), but also protect it (spin protection). We employ two laser
pulses, one pump with field $E_1(t)$ and one dump with $E_2(t)$
\cite{japan}. The laser parameters of the pump are the same as
above. The dump is tuned to the transition between $p_1$ and $d_6$
(see fig. \ref{fig1}(a)), with duration of 60 fs and field amplitude
of 2.5 times the pump. These parameters are carefully chosen to
maximize the effect and are fixed below. The only variable is the time
delay $T=t_1-t_2$ between the pump and dump. Figure \ref{fig2}(d)
shows the populations as a function of $T$ for three states, with the
solid line for the originally occupied $d_5$-state, dashed line for
the originally unoccupied $p_1$-state, and circle line for the
originally unoccupied $d_6$-state. Negative time delay means the dump
proceeds earlier than the pump. {The results in
  figs. \ref{fig2}(d) and 3(e) are computed after $t=200$ fs. If the
  results oscillate with time, we show the time-averaged data.}  At
$T=-60$ fs, similar to the above single-pump-pulse excitation, $d_5$
loses population to $p_1$, but now the {$d_6$-state} gains
some population, in sharp contrast to the above single-pulse
excitation. When $T$ decreases (less negative), the population in
$d_5$ increases sharply, but the populations in $p_1$ and $d_6$ drop
sharply. In other words, the system becomes more difficult to excite
and much less absorptive, or transparent to the light field.  This is
a manifestation of electromagnetically induced transparency (EIT)
\cite{eit}, though in EIT normally cw lasers are used \cite{eit}. What
is novel here is that a time delay can induce EIT, which has a direct
consequence in the spin change. Figure \ref{fig2}(e) illustrates that
during EIT, the spin is mostly unchanged, i.e., {spin
  protection}. What is even more interesting is that when we increase
the time delay over 60 fs, population inversion starts.

Going back to fig. \ref{fig2}(d), we see that the population in $d_6$
becomes largest and that in $d_5$ drops below both $d_6$ and $p_1$, a
total population inversion. Now if we monitor the spin change at the
same delay, we find that $s_z$ changes its sign from positive to
negative (see fig. \ref{fig2}(e)), or spin switching.  We expect our
predictions are detectable experimentally in magnetic systems whose
electronic and magnetic structures are similar to those of our
model. The key to the successful realization of EIT is that the
dipole-transition between the LUMO and HOMO must be forbidden and the
transition between HOMO and LUMO+1 or higher states is allowed.  Very
recently, H\"ubner \et demonstrated the optical spin manipulation for
magnetic logic operations in the $\rm Ni_3Na_2$
cluster \cite{wolfgang09,lefkidis}. Therefore, the possibility to
observe our prediction is very high.

\section{Conclusions}

We have demonstrated a new electromagnetically induced {spin
  protection and switching} in a specially designed {magnetic
  dot}. The minimum requirement for such a dot is that the optical
transition is allowed between the HOMO and LUMO+1, but not HOMO and
LUMO. A strong dump pulse couples LUMO and LUMO+1, and directly
controls the spin dynamics. As a result, the spin becomes harder to
excite, i.e., {spin protection}. We have also shown that if
the time delay between the pump and dump laser pulses exceeds 60 fs, a
spin switch from spin up to spin down occurs. We expect both these
predictions are detectable experimentally.

This work was supported by the U. S. Department of Energy under
Contract No. DE-FG02-06ER46304. We acknowledge part of the work as
done on Indiana State University's high-performance computers, which
is supported by the Center for Instruction, Research and
Teaching. This research used resources of the National Energy Research
Scientific Computing Center at Lawrence Berkeley National Laboratory,
which is supported by the Office of Science of the U.S. Department of
Energy under Contract No. DE-AC02-05CH11231. Initial studies used
resources of the Argonne Leadership Computing Facility at Argonne
National Laboratory, which is supported by the Office of Science of
the U.S. Department of Energy under Contract No. DE-AC02-06CH11357.

\mbox{$^{*}$Electronic address: gpzhang@indstate.edu}

\end{document}